\documentclass[12pt,twoside]{article}
\usepackage{fleqn,espcrc1}
        
\usepackage{graphicx}
\usepackage[figuresright]{rotating}

\title{Non-extensive statistics effects in quark-gluon plasma and 
in relativistic heavy-ion collisions}

\author{W.M. Alberico, \address{Dipartimento di Fisica and INFN, 
Universit\`a di Torino, Italy}  
A. Lavagno \address{Dipartimento di Fisica and INFN, 
Politecnico di Torino, Italy}
and P. Quarati $^{\rm b}$ 
}
       
\begin{document}

\maketitle

\begin{abstract}
The presence of 
memory effects and color long-range forces among the many-parton system in 
the early stage of heavy-ion collisions can affect the particle statistical 
behavior at the freeze-out temperature. 
In this context, we calculate, in the framework of the equilibrium 
generalized non-extensive thermostatistics, 
the shape of pion transverse mass spectrum and the value of  
the transverse momentum correlation function of the pions 
emitted during the central Pb+Pb collisions and we show that the 
experimental results is well reproduced assuming very small deviations 
from the standard statistics.
\end{abstract}

\section{Introduction}

Lattice QCD calculations predict that 
nucleons brought to high enough temperature or pressure 
will make a transition to a 
state where the quarks are no longer confined into individual hadrons but 
dissociate into a plasma of quarks and gluons (QGP). 

Such a QGP can be found in: early universe, dense and hot stars, neutron 
stars and nucleus-nucleus high energy collisions where heavy-ions are 
accelerated to relativistic energies and made to collide.
After collision, a fireball is created, expands, cools, freezes-out into 
hadrons, photons, leptons that are detected and analysed.

Under the hypothesis that QGP is generated in the early stage of the 
relativistic collisions, the quantities characterizing the plasma, such as 
lifetime and damping rate of quasiparticles, are usually calculated within 
finite temperature perturbative QCD 
since the interactions among quarks and gluons 
become weak at small distance 
or high energy. However, this is rigorously true 
only at very high temperature (even up to five times the critical 
temperature) while around the critical 
temperature (that corresponds approximately to the energy scale of the 
SPS-Cern experiments) lattice calculations show that there are strong 
non-perturbative QCD effects and QGP cannot be considered as a weakly 
interacting plasma \cite{scha,ban}.

In order to understand well the relevance of this point, let us consider the 
color-Coulombic QGP parameter, which can be expressed as 
the ratio of the average potential energy to the average kinetic energy 
of particles 
\cite{ban} 
\begin{equation}
\Gamma=\frac{\langle P.E.\rangle}{\langle K.E.\rangle}
\approx \left (\frac{4\pi n}{3}
\right)^{1/3}\frac{4}{3}\frac{\alpha_s}{T} \; ,
\end{equation}
In perturbative QCD, $\alpha_s\rightarrow 0$, $\Gamma\ll 1$ and the plasma 
is considered as an ideal gas.
At $T = T_c\approx 200$ MeV, $\alpha_s=g^2/4\pi \approx 0.5$, 
$<r>\approx 1$ fm and we get $\Gamma\approx 2/3<1$.
Therefore, during  hadronization, the plasma parameter 
lies in an intermediate region typical of a non-ideal plasma: collective and 
individual effects coexist; memory effects and long-range interactions 
are present. On the other hand, 
it is not a true strongly coupled plasma ($\Gamma>1$),  
and a suitable approach describing 
a system at these intermediate conditions is still non available.

Recent progresses in statistical mechanics have shown that 
the non-extensive statistics, proposed by Tsallis \cite{tsa}, 
can be considered as the natural generalization of the extensive 
Boltzmann-Gibbs statistics in presence 
of long-range forces and/or in irreversible processes related to 
microscopic long-time memory effects. Since these features are present 
in the early stage of the collisions, we argue that the non-extensive 
statistics  can be more appropriate than the Boltzmann--Gibbs one in 
the context of high-energy heavy-ion collisions.
Here we investigate how these statistical effects can affect 
the equilibrium properties of experimental observables such as the transverse 
momentum spectrum and the fluctuation-correlation measure of pions 
emitted during the collisions.

\section{Transverse momentum spectrum and $q$-blue shift}

The transverse momentum distribution depends on the phase-space distribution 
and usually an exponential shape is employed to fit the experimental data. 
This shape is obtained by assuming a purely thermal source with a Boltzmann 
distribution. High energy deviations from the exponential shape are taken 
into account by introducing a dynamical effect due to collective transverse 
flow,  also called blue-shift. 

Let us consider a different  point of view and argue that the deviation 
from the Boltzmann slope at high $p_\perp$ can be ascribed to the presence 
of non-extensive statistical effects in the steady state distribution of 
the particle gas. In this framework, at the first order 
in $(q-1)$ the transverse mass spectrum can be written as \cite{alb}

\begin{equation}
\frac{dN}{m_\perp dm_\perp}=C \; m_\perp \left\{ K_1\left ( z \right )+ 
\frac{(q-1)}{8} z^2 \; \left [3 \; K_1 (z)+K_3 (z) \right ]  \right\} \;,
\label{mpt}
\end{equation}
where $z=m_\perp/T$, $m_\perp=\sqrt{p^2+m^2}$, 
$K_1$ and $K_3$ are the modified Bessel function of the first and the third 
order, respectively. 
In the asymptotic limit, $z\gg 1$, we have 
\begin{equation}
\frac{dN}{m_\perp dm_\perp}=D \; \sqrt{m_\perp} \, 
\exp \left (-z+\frac{q-1}{2}\, z^2\right )\; , 
\label{mpst}
\end{equation}
and we may obtain the generalized slope parameter or $q$-blue shift (if $q>1$)
\begin{equation}
T_q=T+ (q-1) \, m_\perp \; .
\label{qtslope}
\end{equation}
Let us notice that the slope parameter depends on the detected particle mass 
and it increases with the energy (if $q>1$) as it was observed in the
experimental results \cite{na44}.

\begin{figure}[htb]
\begin{minipage}[t]{80mm}
\includegraphics[scale=0.6]{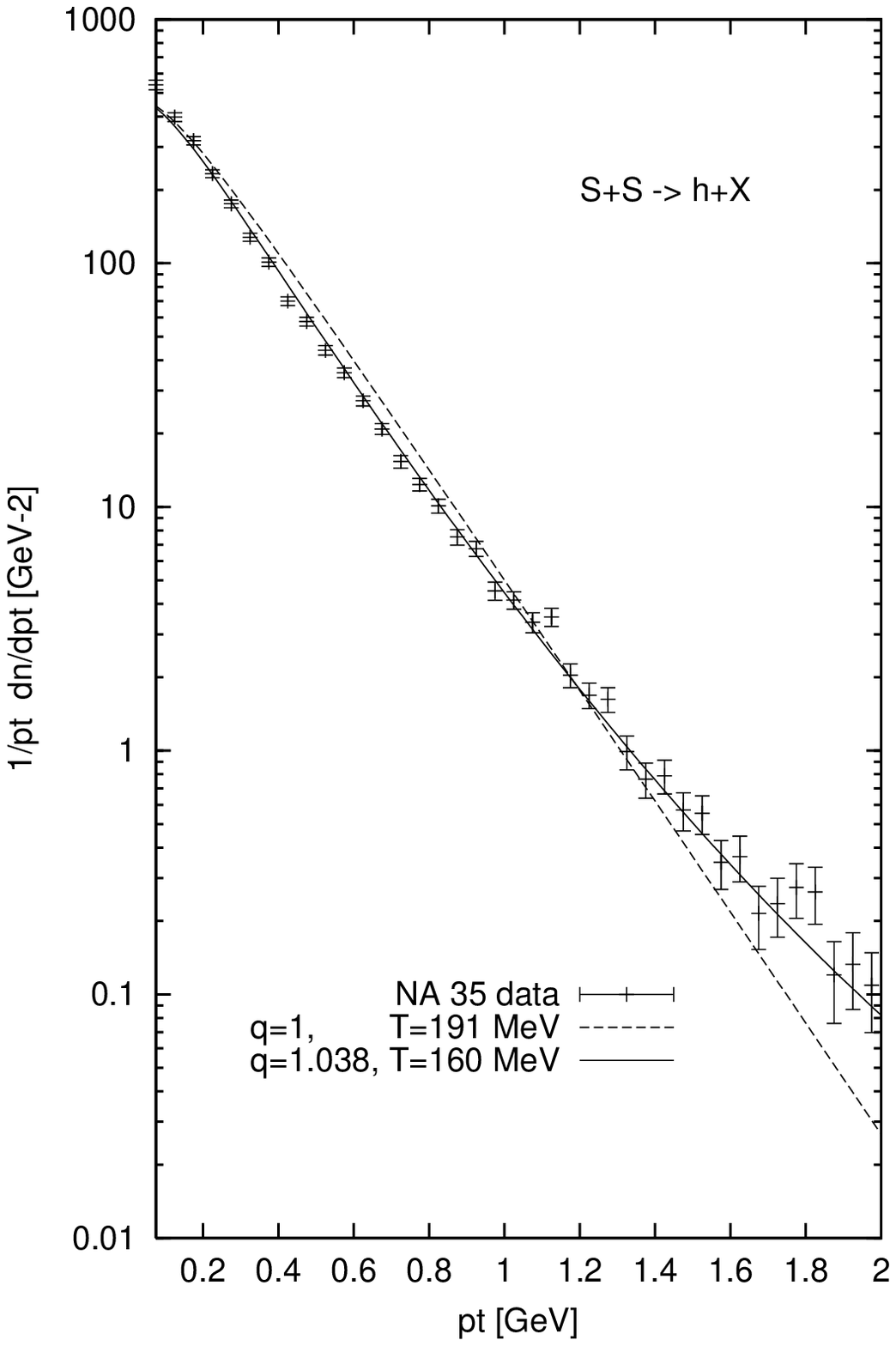}
\vspace{-1.0cm}
\caption{Experimental NA35 \cite{na35} 
transverse momentum distribution compared with the  
exponential ($q=1$) thermal distribution (dashed line) and the 
modified thermal distribution shape (solid line) 
by using non-extensive statistics ($q=1.038$)}
\label{sspt}
\end{minipage}
\hspace{\fill}
\begin{minipage}[t]{75mm}
\includegraphics[scale=0.6]{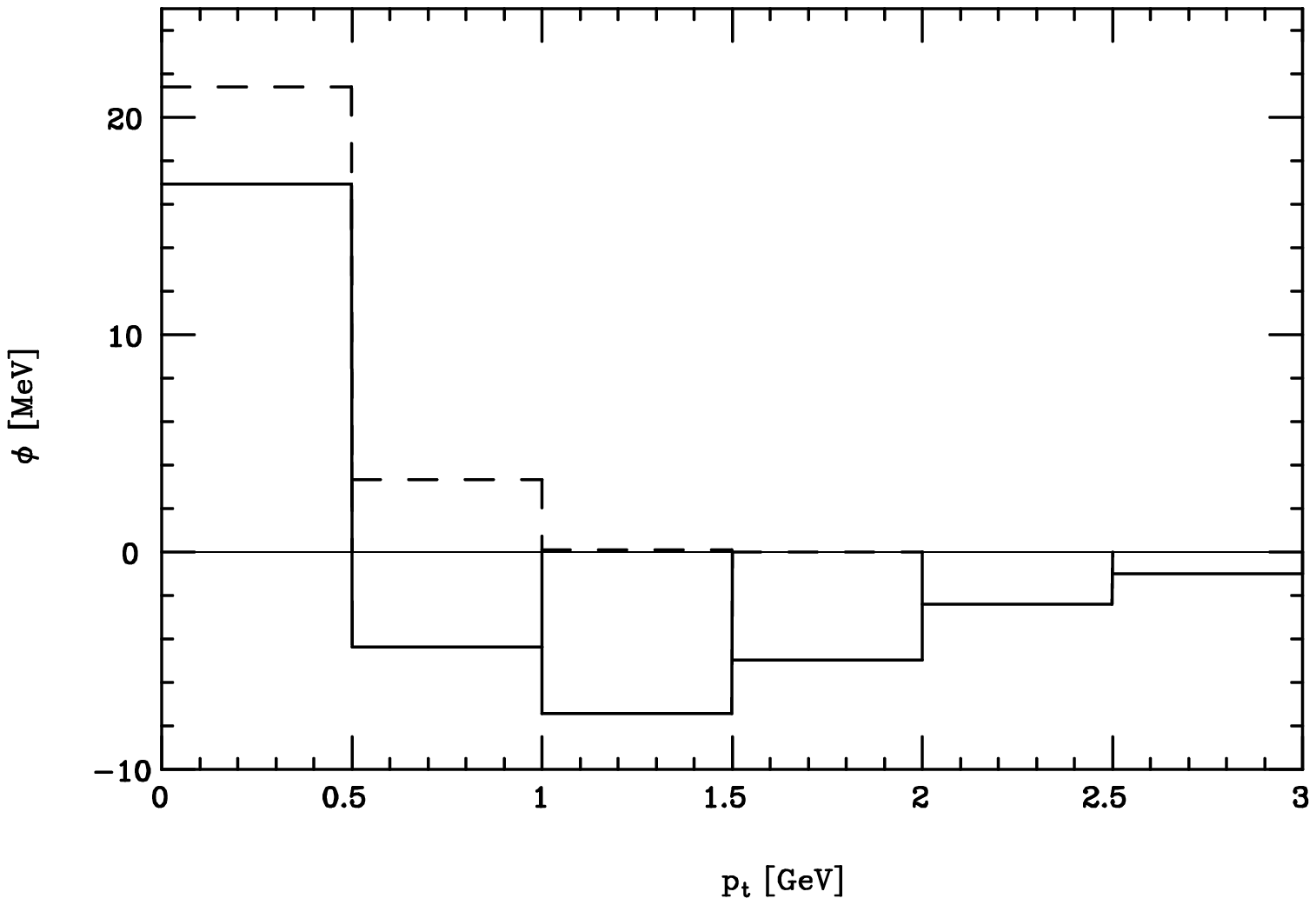}
\vspace{-1.0cm}
\caption{The partial contributions to the correlation measure 
$\Phi_{p_\perp}$ [MeV] in different $p_\perp$ intervals, 
at $T=170$ MeV and $\mu=60$ MeV. The dashed 
line refers to standard statistical calculations with $q=1$, the
solid line corresponds to $q=1.038$.}
\label{isto}
\end{minipage}
\end{figure}

In Fig. \ref{sspt}, we report the experimental $S+S$ transverse momentum 
distribution (NA35 data \cite{na35}) 
compared with the purely thermal distribution ($q=1$) 
and the one obtained in the framework of Tsallis statistics ($q=1.038$). 
We will use, consistently, the same value of $q$ 
to determine the experimental transverse momenta fluctuations. 
Similar calculations in the framework of 
non--extensive statistics have been done in Ref.\cite{wilk}.

\section{Transverse momentum fluctuations}

Ga\'zdzicki and Mr\'owczy\'nski 
introduced the following quantity \cite{gaz92,mro} 
\begin{equation}
\Phi_{p_\perp}=\sqrt{\langle Z_{p_\perp}^2 \rangle \over \langle N \rangle} -
\sqrt{\overline{z_{p_\perp}^2}} \;,
\label{phix}
\end{equation}
where 
$z_{p_\perp}= p_\perp - \overline{p_\perp}$
and $Z_{p_\perp} = \sum_{i=1}^{N}(p_{\perp i} - \overline{p}_{\perp i})$, 
$N$ is the multiplicity of particles produced in a single event.
Non-vanishing $\Phi$ implies effective correlations among particles which
alter the momentum distribution.

In the framework of non--extensive statistics and 
keeping in mind that it preserves 
the whole mathematical structure of the thermodynamical relations,
 it is easy to show that the 
two terms in the right hand side of Eq.(\ref{phix}) can be expressed in 
the following simple form
\begin{eqnarray}
\overline{z^2_{p_{\perp}}} =
{1 \over \rho}\int{d^3p \over (2\pi )^3} \, \, \Big(p_{\perp} - 
\overline{p}_{\perp} \Big)^2   \langle n \rangle_q \; ,
\ \ \  {\rm and} \ \ \  
{\langle Z_{p_{\perp}}^2 \rangle \over \langle N \rangle }=
{1 \over \rho}\int{d^3p \over (2\pi )^3}
\,\Big(p_{\perp} - \overline{p}_{\perp} \Big)^2   
\langle\Delta n^2\rangle_q \;,
\label{qz2p}
\end{eqnarray}
where 
\begin{equation}
\overline{p}_{\perp} = {1 \over \rho}\int{d^3p \over (2\pi )^3} \;
p_{\perp} \langle n \rangle_q \ \ \ \ \ \ \  {\rm with} 
\ \ \ \ \ \ \ \rho=\int{d^3p \over (2\pi )^3} \langle n \rangle_q \;.
\end{equation}

In the above equations we have indicated with $\langle n \rangle_q$ the
following mean 
occupation number of bosons (valid only for dilute gas and/or 
small value of $q$, see Ref.\cite{buyu} for details)
\begin{equation}
\langle n\rangle_q=\frac{1}{[1+(q-1)\beta (E-\mu)]^{1/(q-1)}\pm 1} \;,
\label{distri}
\end{equation}
and with 
$\langle\Delta n^2\rangle_q=\langle n^2\rangle_q-\langle n \rangle^2_q$ the 
generalized particle fluctuations, given by 
\begin{eqnarray}
\langle\Delta n^2\rangle_q\equiv
\frac{1}{\beta}\frac{\partial\langle n\rangle_q}{\partial\mu}=
\frac{\langle n\rangle_q }{1+(q-1)\beta (E-\mu)}\, 
(1\mp \langle n\rangle_q)
=\langle n\rangle_q^q \; (1 \mp \langle n\rangle_q )^{2-q}.
\label{fluc}
\end{eqnarray}

NA49 Collaboration has recently measured the correlation 
$\Phi_{p_\perp}$ of the pion transverse momentum (Pb+Pb at 
158 A GeV) \cite{na49} obtaining  
$\Phi^{exp}_{p_\perp}=(0.6\pm 1) \; {\rm MeV}$. 
This value is the sum of two contributions: 
$\Phi^{st}_{p_\perp}=(5\pm 1.5) \; {\rm MeV}$, 
the measure of the statistical two-particle correlation, and 
$\Phi^{tt}_{p_\perp}=(-4\pm 0.5) \; {\rm MeV}$, 
the anti-correlation from limitation in two-track resolution.    

Standard statistical calculations ($q=1$) give \cite{mro}
$\Phi^{st}_{p_\perp} = 24.7\; {\rm MeV \; at \; } T = 170 \; {\rm MeV}, 
\ \ \mu = 60 \; {\rm MeV}$. 
In the frame of non--extensive statistics, instead, for $q=1.038$, we obtain
the experimental (statistical) value: 
$\Phi^{st}_{p_\perp} = 5 \; {\rm MeV \; at\;} T = 170 \; {\rm MeV}, 
\ \ \mu = 60 \; {\rm MeV}$. 

In Fig. \ref{isto}, we show the partial contributions to the 
quantity $\Phi_{p_\perp}$, by using Eq.s (\ref{qz2p}), 
and by extending the integration over $p_\perp$ to partial
intervals $\Delta p_\perp=0.5$~GeV at $T=170$ MeV and $\mu=60$ MeV.   
In the standard statistics (dashed line), 
$\Phi_{p_\perp}$ is always positive and vanishes in the $p_\perp$-intervals 
above $\approx 1$ GeV. In the non-extensive statistics (solid line), instead, 
the fluctuation measure  $\Phi_{p_\perp}$ 
becomes negative for $p_\perp$ larger than $0.5$ GeV 
and becomes vanishingly small only in $p_\perp$-intervals above $\sim 3$ GeV.
Such a negative value of $\Phi_{p_\perp}$ at high $p_\perp$ could be 
a clear and unambiguos evidence of the presence of non-extensive regime in 
heavy-ions collisions.

{}

\end{document}